\documentclass[12pt]{iopartm}

\usepackage{graphicx}

\newcommand{\ind}[2]{^{#1}_{\mbox{\scriptsize #2}}}
\newcommand{\al}[2]{\alpha\ind{#1}{#2}}
\newcommand{\aktw}{\alpha\ind{}{{\tiny KTW}}}
\newcommand{\ro}[1]{\rho^{(#1)}}
\newcommand{\sla}{\slash \hspace{-0.22cm}}
\newcommand{\slaq}{\slash \hspace{-0.18cm}}      

\def\fpb{\frac{4 \pi}{\bz}}
\def\nf{n_{\mbox{\scriptsize f}}}
\def\bz{\beta_0}
\def\fpi{f_{\pi}}
\def\KL{K\"all\'en--Lehmann }
\def\SD{Schwinger--Dyson }

\begin{document}


\title[]{Infrared enhanced analytic coupling and \\
chiral symmetry breaking in QCD}

\author{A C Aguilar$^{1}$, A V Nesterenko$^{1,2}$ and 
J Papavassiliou$^{1}$}


\address{$^{1}$ Departamento de F\'\i sica Te\'orica
and IFIC, Centro Mixto, \\ Universidad de Valencia--CSIC,
E-46100, Burjassot, Valencia, Spain}

\address{$^{2}$ Bogoliubov Laboratory of Theoretical Physics, \\
Joint Institute for Nuclear Research,
Dubna, 141980, Russian Federation}

\ead{Joannis.Papavassiliou@uv.es}


\begin{abstract} 
We study the impact on chiral symmetry breaking of a recently
developed model for the QCD analytic invariant charge. This charge
contains no adjustable parameters, other than the QCD mass scale
$\Lambda$, and embodies asymptotic freedom and infrared enhancement
into a single expression. Its incorporation into the standard form of
the quark gap equation gives rise to solutions for the dynamically
generated mass that display a singular confining behaviour at the
origin. Using the Pagels--Stokar method we relate the obtained
solutions to the pion decay constant $\fpi$, and estimate the scale
parameter $\Lambda$, in the presence of four active quarks, to be
about $880\,$MeV.
\end{abstract}



\maketitle

\setcounter{footnote}{0}
\def\fnsymbol{\arabic}

\section{Introduction}
\label{Sect:Intro}

     The lack of a systematic calculational scheme applicable to  the
infrared sector of Quantum Chromodynamics (QCD) has motivated the
advent of models  aspiring to provide  phenomenologically viable 
links between asymptotic freedom and confinement, by   enriching
perturbation theory  with judiciously selected nonperturbative
information.  An important source of such information is provided by
the dispersion relations. The latter, being based on the ``first
principles'' of the theory, furnish  the definite analytic properties
with respect to a given kinematic variable of a physical quantity
under consideration.  The basic idea behind the so-called  ``analytic
approach'' to Quantum Field Theory (QFT) is to supplement 
perturbation theory, and in particular the renormalization group (RG)
formalism, with the nonperturbative information encoded in the
relevant dispersion relations. Specifically, the perturbative
solutions of the RG equations  possess unphysical singularities in
the infrared domain,  a fact that contradicts the general principles
of local~QFT. The analytization procedure amounts to the restoration
of the correct analytic  properties for a physical quantity at hand,
by forcing it to satisfy the \KL spectral representation (see
equation~(\ref{DefAn})). This method was first proposed in the
framework of Quantum Electrodynamics (QED) and applied to the study
of the invariant charge of the theory~\cite{AQED}. Later on, these
considerations were generalized to the case of the non-Abelian
theories, and the analytic approach to QCD~\cite{ShSol} emerged. This
approach has been successfully applied to the study of the strong
running coupling~\cite{ShSol,PRD}, perturbative series for the QCD
observables~\cite{APTRev}, and some intrinsically nonperturbative
aspects of the hadron dynamics~\cite{PRD,Review}. Some of the main
advantages of the analytic approach to QCD are the absence of
unphysical singularities (by construction), and a fairly good
higher-loop and scheme stability of the results obtained.

     A central quantity within the aforementioned approach  is
$\alpha(k^2)$, the running  (or ``effective'') coupling (charge) of 
QCD. Clearly, this quantity  can  be reliably computed only  in  
the  ultraviolet  region,  where   perturbation  theory  is
applicable,  and  must  be  modelled  in  the  infrared domain, 
where perturbative methods break down, and one eventually encounters
the unphysical singularities, such as the Landau pole. The analytic 
approach to QCD eliminates such artefacts and provides  a concrete
analytic expression for the  running coupling in the infrared, 
maintaining at the same time the standard asymptotic behaviour  in
the ultraviolet. However, since the analyticity requirement can be
incorporated into the RG formalism in various ways, two main pictures
have emerged.  Specifically, if the analyticity condition is  imposed
directly on the perturbative running coupling $\al{}{s}(k^2)$ one
arrives at the model of~\cite{ShSol} with finite universal infrared
limiting value given by $\alpha(0)=4\pi/\beta_0$. If, instead, the
analyticity requirement is imposed on the corresponding $\beta$
function, one obtains the running coupling of~\cite{PRD} which is
singular  (``enhanced'') at~$k^2=0$. The latter invariant charge has
proved to  be able to describe a number of the strong interaction
processes both of perturbative and of nonperturbative
nature~\cite{PRD,Review}. Undoubtedly, it would be interesting to
further study the physical  implications and phenomenological
possibilities offered by the infrared enhanced analytic invariant
charge.  The primary objective of this paper is to examine the 
compatibility of the infrared enhanced analytic running coupling with
chiral symmetry breaking~(CSB), and its impact on the solutions  of
the \SD (SD) equation controlling the dynamical generation of  quark
masses. 

     The way the running coupling enters into the standard SD
equation for the quark propagator (``gap equation'') is rather 
well-known.  Since QCD is not a fixed point theory, the QED--inspired
gap equation must be modified, in order to incorporate  the
asymptotic freedom. The usual way of accomplishing this eventually
boils down to the replacement $1/k^2 \to \alpha(k^2)/k^2$ in the
corresponding kernel of the gap equation. The inclusion of
$\alpha(k^2)$ is essential for arriving at an integral equation for
the quark propagator~$S(p)$ which is well-behaved in the ultraviolet.
Indeed, the additional logarithm in the denominator of the kernel due
to the  strong running coupling $\alpha(k^2)$ improves the
convergence of the integral. However, since the perturbative form of
$\alpha(k^2)$ diverges  as $1/\ln(k^2/\Lambda^2)$ when $k^2 \to
\Lambda^2$, a physically motivated  model for the infrared behaviour
of  $\alpha(k^2)$ is needed. The infrared enhanced analytic charge
represents a good candidate for such a purpose, since, as has been
explained in detail in~\cite{PRD,Review}, it combines asymptotic
freedom and  confinement behaviour in a single expression.  This is
to be contrasted with the standard  method of introducing  asymptotic
freedom and confinement (see, e.g.,~\cite{VScheme}),  whereby the
running coupling  is composed by two separate, and theoretically
rather uncorrelated,  contributions (see further discussion in
Section~\ref{Sect:Asympt}).

     The paper is organized as follows. In Section~\ref{Sect:AIC}  we
briefly review the  most salient features of the analytic approach,
with particular emphasis on the running coupling displaying infrared
enhancement. In Section~\ref{Sect:GAP} we go over the usual
assumptions and approximations entering into the derivation of the
standard gap equation. In Section~\ref{Sect:Asympt}  we study the
asymptotic behaviour of the dynamical quark mass function, by
converting the integral  equation into a differential form. In
Section~\ref{Sect:NumSol} the integral equation is solved
numerically, and the results are presented. It turns out that,  with
four active quarks, the experimental value of the pion decay constant
$\fpi$  may be obtained if the QCD mass-scale $\Lambda$, which  is
the only free parameter available, acquires the  rather elevated
value of about $880\,$MeV.  Finally, in Section~\ref{Sect:Concl} we 
discuss the results and present our conclusions.

\section{Analytic invariant charge in QCD}
\label{Sect:AIC}

     In this Section we give a brief summary of one of the models 
for the QCD analytic invariant charge~\cite{PRD}. This model shares
all the advantages of the analytic approach and it  was successful in
the description of a wide range of QCD phenomena both of perturbative
and of intrinsically nonperturbative nature~\cite{Review}.
Furthermore, it is of a particular interest to note that this model
has recently been re-derived~\cite{Schrempp}, proceeding from
completely different motivations.

     As has already been commented in the Introduction,  the basic
idea behind the analytic approach to QFT~\cite{AQED,ShSol} is to
supplement  perturbation theory  with the nonperturbative information
encoded in the relevant dispersion relations. The latter are based on
``first principles'' of the theory, and provide one with the definite
analytic properties of a  given  physical quantity  with respect to a
proper kinematic variable. In practice the ``analytization
procedure''~\cite{ShSol} amounts to  the restoration of the correct
analytic properties for a given quantity $F(k^2)$ by imposing the \KL
integral representation\footnote{A metric with signature $(-1, 1, 1,
1)$ is used, so that positive $k^2$ corresponds to a spacelike
momentum transfer.}
\begin{equation}
\label{DefAn}
\Bigl\{F(k^2)\Bigr\}_{\mbox{$\!$\scriptsize an}} =
\int_{0}^{\infty} \!\frac{\varrho(\sigma)}{\sigma+k^2}\,
d \sigma.
\end{equation}
Here the spectral function $\varrho(\sigma)$ is determined by the
initial (perturbative) expression 
\begin{equation}
\varrho(\sigma) = \frac{1}{2\pi i}\, \lim_{\varepsilon \to 0_{+}}
\Bigl[F(-\sigma-i\varepsilon)-F(-\sigma+i\varepsilon)\Bigr],
\end{equation}
with $\sigma>0$.
   
     The model for the analytic invariant charge~\cite{PRD,Review}
that we will study in this paper is obtained by imposing the
requirement of analyticity~(\ref{DefAn}) on the perturbative
expansion of the RG $\beta$~function
\begin{equation}
\label{RGEqAn}
\frac{d\,\ln \al{(\ell)}{an}(\mu^2)}{d\,\ln \mu^2} =
- \left\{\sum_{j=0}^{\ell-1} \beta_{j}
\left[\frac{\al{(\ell)}{s}(\mu^2)}{4 \pi}
\right]^{j+1}\right\}\ind{}{$\!$an}.
\end{equation}
In this equation  $\al{(\ell)}{an}(\mu^2)$ is the $\ell$-loop
analytic  invariant charge, $\al{(\ell)}{s}(\mu^2)$ denotes the 
$\ell$-loop perturbative  running coupling, $\beta_{j}$ stands for
the $\beta$~function expansion coefficient ($\beta_{0} = 11 - 2 \nf /
3,\,$  $\beta_{1}=102 - 38 \nf / 3, \mbox{...}\,$), and $\nf$ is the 
number of active quarks.  It is worth noting that
prescription~(\ref{RGEqAn}) differs from that of  the original
Shirkov--Solovtsov model~\cite{ShSol}, where the 
requirement~(\ref{DefAn}) was imposed directly on the perturbative
running coupling~$\al{}{s}(k^2)$ (discussion of this issue can also 
be found in~\cite{PRD,Review,MPLA,DV02}).

     At the one-loop level the RG equation~(\ref{RGEqAn}) can be
solved explicitly~\cite{PRD}:
\begin{equation}
\label{AIC1L}
\al{(1)}{an}(k^2) = \fpb\, \frac{z - 1}{z \, \ln z},
\qquad z = \frac{k^2}{\Lambda^2}.
\end{equation}
The solution to equation~(\ref{RGEqAn}) can also be represented in
the form of the \KL integral
\begin{equation}
\label{AICHLKL}
\al{(\ell)}{an}(k^2) = \fpb \int_{0}^{\infty}
\frac{\ro{\ell}(\sigma)}{\sigma + z}\, d \sigma,
\end{equation}
where the one-loop spectral density is
\begin{equation}
\label{SpDns1L}
\ro{1}(\sigma) =
\left(1+\frac{1}{\sigma}\right)\frac{1}{\ln^2\!\sigma+\pi^2},
\end{equation}
and the explicit expression for the $\ell$-loop $\ro{\ell}(\sigma)$
can be found in~\cite{PRD,Review}.

\begin{figure}[ht]
\begin{center}
\includegraphics[width=105mm]{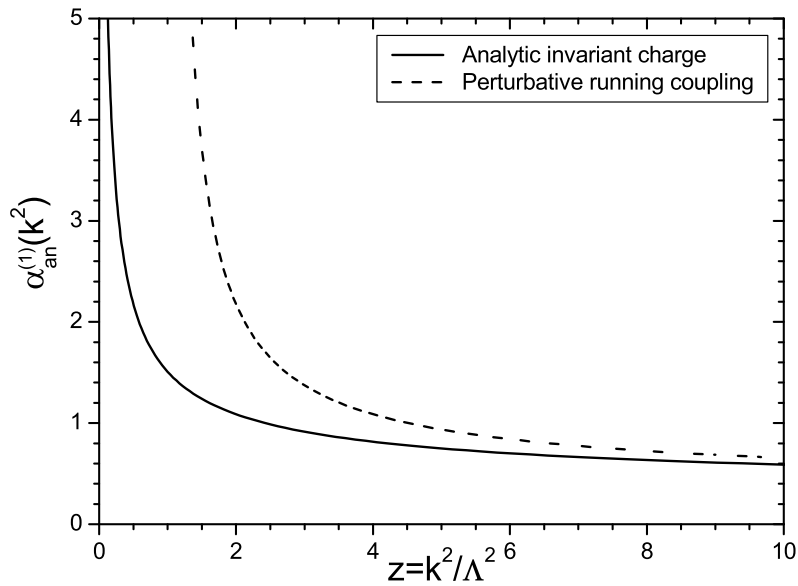}
\end{center}
\caption{The analytic invariant charge $\al{(1)}{an}(k^2)$ (see 
equation~(\protect\ref{AIC1L})) and the perturbative running coupling
$\al{(1)}{s}(k^2)$ at the one-loop level (solid and dashed  curves,
respectively). Here $\nf=4$ active quarks are assumed.}
\label{Plot:AIC}
\end{figure}

     The invariant charge~(\ref{AIC1L}) possesses a number of
appealing features. First of all, it has  the correct analytic
properties in the $k^2$~variable, demanded in equation~(\ref{DefAn}),
namely, it has the only cut $k^2 \le 0$ along the negative semiaxis
of real~$k^2$. Then, it contains no adjustable 
parameters\footnote{It should be noted that the Shirkov--Solovtsov
running coupling~\cite{ShSol} has no adjustable parameters, either.
So, both these models are the ``minimal'' ones in this sense.}. Thus,
similarly to the perturbative approach, the QCD scale
parameter~$\Lambda$ remains the basic characterizing quantity of the
theory. In addition, the invariant charge~(\ref{AIC1L}) incorporates
the ultraviolet asymptotic freedom with the infrared enhancement in a
single expression (see Figure~\ref{Plot:AIC}), that plays an
essential role in applications of the model in hand to the
description of the quenched lattice simulation data 
(see~\cite{Review,Schrempp} for the details). It is worth mentioning
here that the singular behaviour of the strong running coupling
$\alpha(k^2)$ when $k^2 \to 0$ is also  supported by a number of
studies of the SD equations (see, e.g., papers~\cite{KTW, BBZ,
AlekArbu, Gogohia:1993vc} and references therein). Moreover, the
invariant charge~(\ref{AICHLKL}) displays a good  higher loop and
scheme stability, and it has proved to describe a number of strong
interaction processes (e.g., confining static quark--antiquark
potential, inclusive $\tau$~lepton decay) in a self--consistent way.
The detailed analysis of the properties of the analytic running
coupling~(\ref{AICHLKL}) and its applications can be found
in~\cite{PRD,Review,MPLA}.

     Given the characteristic features of the  analytic invariant
charge  mentioned above, it would be interesting to study its
influence on  phenomena particularly sensitive to the low-energy
dynamics. To that end, in the following three sections  we study how
the analytic invariant charge~(\ref{AICHLKL}) affects the  the
mechanism of CSB and dynamical mass generation for the quarks,  
through the study of the SD equations  governing the dynamics of the
quark propagator~\cite{miran, higa, pagels, cjt, roberts, alkofer,
Sauli, Kekez:2004tm, Hashimoto:2002px}. 

\vfill 

\section{The gap equation}
\label{Sect:GAP}

     Throughout this Section, we will work exclusively in Euclidean
space. According to the usual conventions, the starting point is to
express the fully dressed quark propagator in the following general
form~\cite{roberts}:
\begin{equation}
S^{-1}(p) = i\sla{p} +m_0 + \Sigma(p) =
i\sla{p}A(p^2) + B(p^2),
\label{qprop}
\end{equation}
where $m_0$ is the bare quark mass and $\Sigma(p)$ is the quark
self-energy.

     Before proceeding, it is worthwhile to make some additional
comments on the above functions. Since we will be concentrated in the
case where one does not have explicit CSB, i.e., the bare mass
$m_0=0$, the quark mass is generated only through dynamical effects.
In this case, CSB takes place when the self-energy $\Sigma(p)$
develops a nonzero value. Alternatively, one may define the quark
mass function $M(p^2)$ in terms of the functions $A(p^2)$ and
$B(p^2)$, as $M(p^2)=B(p^2)/A(p^2)$; then CSB is considered to occur
when $B(p^2)\neq 0$.

\begin{figure}[ht]
\begin{center}
\includegraphics[width=95mm]{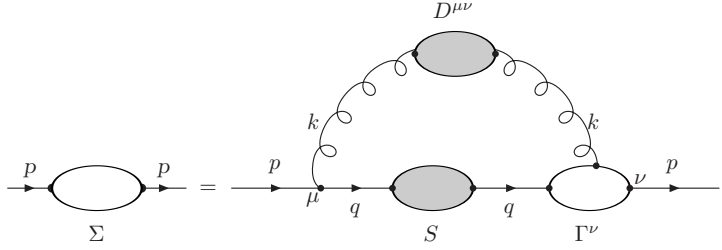} 
\end{center}
\caption{The SD equation (\protect\ref{senergy}) for the quark
self-energy. The black blobs represent the fully dressed quark and
gluon  propagators  and the white one is the proper quark-gluon
vertex.} 
\label{self}
\end{figure}

     The quark SD equation is represented schematically in
Figure~\ref{self} and can be written as
\begin{equation}
\label{senergy}
\Sigma(p)=\frac{4}{3}g^2\int\frac{d^4q}{(2\pi)^4}
\gamma_{\mu}S(q)\Gamma_{\nu}(q,p)D^{\mu\nu}(k),
\end{equation}
where we have used that  $\sum_a \lambda^a \lambda^a=4/3$, 
$\lambda^a$ being the Gell-Mann matrices, and $k=p-q$.  According to
this equation, the self-energy $\Sigma(p)$ is dynamically determined 
in terms of itself, the full   gluon propagator, denoted by 
$D_{\mu\nu}^{ab}(k)=\delta^{ab}D_{\mu\nu}(k)$, and the full
quark-gluon vertex~$\Gamma_{\nu}(q,p)$. Of course, both
$D_{\mu\nu}^{ab}(k)$ and $\Gamma_{\nu}(q,p)$ obey their own
complicated SD equations, a fact which eventually makes the use of
simplifications unavoidable and further modelling of the unknown
functions involved.

     A common approximation employed in the literature (see, 
e.g.,~\cite{KTW,roberts,atk}) is to neglect the ghost contributions
in the quark SD equation, whose effects are supposed to be  partially
accounted for by the fully dressed gluon propagator and the full
quark-gluon vertex. This assumption leads to  a theory with 
Abelian-like characteristics,  where the usual non-Abelian
Slavnov--Taylor identities are replaced by QED-like Ward
identities~\cite{atk}. In particular, for the quark-gluon vertex we
have
\begin{equation}
i\, k^{\mu} \Gamma_{\mu}(p,q)=S^{-1}(p)-S^{-1}(q).
\label{ward}
\end{equation}
The imposition of this so-called ``Abelian approximation'' gives
rise  to further simplifications. Specifically, due to the  (assumed)
validity of equation~(\ref{ward}), the usual QED identity $Z_1=Z_2$,
where  $Z_1$ and $Z_2$ are the renormalization constants  for the
fermion-boson vertex and fermion wave function, respectively,  is
restored. This, in turn, allows the definition of a RG invariant 
quantity, to be denoted by $\alpha(k^2)$, which is the exact analogue
of the  QED effective charge, namely   
\begin{equation}
g^2D_{\mu\nu}(k)=\left\{\delta_{\mu\nu} -
\frac{k_{\mu}k_{\nu}}{k^2}\right\} \frac{4\pi\alpha(k^2)}{k^2}.
\label{prop}
\end{equation}
The exact transversality of the right hand-side of
equation~(\ref{prop}) is the result of working in  the Landau gauge.
A different choice of gauge would have resulted in an additional
longitudinal term of the form  $\xi  k_{\mu} k_{\nu}/k^4$, where
$\xi$ is the  usual gauge-fixing parameter ($\xi=0$ corresponds to
the  Landau gauge, $\xi=1$ to the Feynman gauge). Note that the
nature of  this  extra term is purely tree-level, i.e., there is no  
higher-order dressing involved.

     Furthermore, the Ward-identity~(\ref{ward}) motivates the use of
the time-honored ``Gauge Technique''~\cite{delbourgo}. Specifically,
a nonperturbative Ansatz for the vertex $\Gamma_{\mu}(p,q)$ in terms
of $S(p)$ is postulated, based on the requirement that  it should
satisfy, by construction, equation~(\ref{ward}).  Evidently, such a
construction leaves the  transverse part of $\Gamma_{\mu}(p,q)$
unspecified; the usual  argument around this ambiguity is that,  in a
theory with a mass-gap, the transverse parts of the vertex are
sub-leading in the infrared, and have little or no consequence on 
CSB (see, e.g.,~\cite{delbourgo}). In the rest of our analysis we
will use for  $\Gamma_{\mu}(p,q)$ the simple Ansatz proposed
in~\cite{KTW,atk}
\begin{equation}
i\Gamma_{\mu}(p,q)= iA(p^2)\gamma_{\mu} +
\frac{k_{\mu}}{k^2}\left\{i\left[A(p^2)-A(q^2)\right]\slaq{q} + 
\left[B(p^2)-B(q^2)\right]\right\}. 
\label{vert}
\end{equation}
Choosing the Landau gauge leads to the further simplification
\begin{equation}
g^2D^{\mu\nu}(k)\Gamma_{\nu}(q,p) = 
g^2D^{\mu\nu}(k)A(q^2)\gamma_{\nu},
\label{prod}
\end{equation}
since, in that case, the gluon propagator is completely transverse. 

     Substituting equations~(\ref{qprop}) and~(\ref{prod}) into quark
gap  equation~(\ref{senergy}), one arrives at the commonly used 
coupled system for the quark self-energy \cite{KTW, roberts, natale}
\begin{equation}
\label{aprop}
\fl \left[A(p^2)-1\right]p^2 =
\frac{4}{3}\int\frac{d^4q}{(2\pi)^4}\frac{ 4\pi\alpha(k^2)}{k^2}
\left(p\!\cdot\!q + 2\,\frac{p\!\cdot\!k \; q\!\cdot\!k}{k^2}\right) 
\frac{A^2(q^2)}{q^2A^2(q^2)+B^2(q^2)}
\end{equation}
and
\begin{equation}
\label{bprop}
\fl B(p^2) = 4\int\frac{d^4q}{(2\pi)^4}\frac{4\pi\alpha(k^2)}{k^2}
\frac{A(q^2)B(q^2)}{q^2A^2(q^2)+B^2(q^2)}.
\end{equation}
Note that the angular integration can be easily evaluated in the last
two equations, by resorting to the usual angle approximation  for
the running coupling~\cite{KTW, atk, aguilar}
\begin{equation}
\label{angle}
\alpha\left((p-q)^2\right) \approx
\theta(p^2-q^2)\alpha(p^2) + \theta(q^2-p^2)\alpha(q^2).
\end{equation}
In particular, with such an approximation, the angular integral for
the Dirac-vector component of the quark self-energy vanishes, leading
automatically to $A(p^2)=1$. Therefore, we can straightforwardly 
relate $B(p^2)$ to the dynamical mass $M(p^2)$.  Then, the coupled
system formed by  equations~(\ref{aprop}) and (\ref{bprop}), reduces
to one single equation, namely~\cite{atk}
\begin{equation}
\label{sde}
{\mathcal M}(x) =\frac{1}{\pi}\left[
\frac{\alpha(x)}{x}\int_0^x \frac{y {\mathcal M}(y)}
{y+{\mathcal M}^2(y)} dy + 
\int_x^{\infty} \frac{\alpha(y) {\mathcal M}(y)}
{y+{\mathcal M}^2(y)} dy
\right],
\end{equation}
where ${\mathcal M}(x)= M(x\Lambda^2)/\Lambda$, $\,x=p^2/\Lambda^2$,
and $y=q^2/\Lambda^2$.

\section{Asymptotical behaviour of the mass function}
\label{Sect:Asympt}

     The main effect of implementing the substitution described by 
equation~(\ref{prop}) at the level of the  gap equation is to
transfer  our ignorance regarding the   behaviour of the gluon
propagator into the nonperturbative  structure of the invariant
charge $\alpha(k^2)$.  This latter quantity can be modelled more
directly, essentially  because it enters more naturally than the
gluon propagator in the parametrization of the low-energy QCD data.
In the following sections, we will study in detail  the solutions
obtained from the quark gap equation~(\ref{sde}) after using as
$\alpha(k^2)$  the one-loop  analytic running coupling
$\al{(1)}{an}(k^2)$~(\ref{AIC1L}). 

     The inclusion or not of  appropriately modelled confinement
effects has an important impact  on the type of solutions that one
obtains from the gap equation.  In fact, it has often be claimed in
the literature that, if such effects  are not included,  one may not 
encounter non-trivial solutions to the gap equation at all, i.e., CSB
does not occur (see, e.g.,~\cite{Papavassiliou:1991hx}). The usual
way of accounting for confinement effects at the  level of the gap
equation is to insert a gluon propagator of the form $\Lambda^2/k^4$,
the (spacial) Fourier transform of linearly rising quark-antiquark
potential \cite{west}. Evidently, this expression fails to capture
asymptotic freedom in the ultraviolet domain, whose effects are
separately supplemented through the inclusion of the corresponding 
perturbative contributions. Thus, the ``standard'' way of describing
both effects is through a running coupling of the form 
\begin{equation}
\label{KTW}
\aktw(k^2) = \frac{C}{z} + \fpb \frac{1}{\ln(z + \tau)}, \qquad 
z=\frac{k^2}{\Lambda^2},
\end{equation}
where $C$ and $\tau$ are dimensionless constants (see,
e.g.,~\cite{KTW}). Usually, $C$ is treated as an adjustable 
parameter, to be determined in such a way as to  reproduce the
correct phenomenological values for the quark masses,  pion decay
constant, and chiral condensates. On the other hand, $\tau$ plays the
role of an infrared regulator; it is  often  treated as an arbitrary
parameter, but in a more complete, physically motivated picture of
QCD, it has been identified   as a dynamically generated gluon
``mass''~\cite{Papavassiliou:1991hx, JMC82, Cornwall:1989gv}.  Note 
that, if $C=0$, the logarithmic  term on the right hand-side of 
(\ref{KTW}) must overcome comfortably  an infrared critical coupling
of about 1.2, in order to obtain from the gap equations
phenomenologically interesting  solutions. This may or may not be
possible, depending on whether $\tau$ is considered as a free
parameter, or if some physical arguments  constrain its possible
range of values, as is the case with the gluon~``mass''.

     A definite advantage of the analytic invariant charge
$\al{(1)}{an}(k^2)$ (\ref{AIC1L}), compared to (\ref{KTW}), is  the
simultaneous incorporation of asymptotic freedom  and infrared
enhancement into a single expression, of concrete theoretical origin,
namely the analyticity properties of the theory. In particular, note
that, unlike a coupling of the type given in~(\ref{KTW}), the running
coupling~(\ref{AIC1L}) contains  no adjustable parameters. This
theoretically appealing feature, is, of course, much more restrictive
when one handles phenomenologically  relevant quantities. It is worth
noting also that the enhancement displayed by the analytic coupling 
$\al{(1)}{an}(k^2)$ has been shown to correspond to the confining
static quark-antiquark  potential with a quasilinear rising behaviour
at large distances. Namely, $V(r)\simeq r/\ln r$, when $r \to
\infty$, with $r$ being the dimensionless distance between quark and
antiquark, see~\cite{PRD,Review} for the details.

     Before proceeding with the numerical solution for the SD
equation~(\ref{sde}) for all range of momenta, it would be
interesting  to gain some explicit insight into its behaviour in the
deep  infrared region. On general grounds, given that the running
coupling (\ref{AIC1L}) diverges at the origin, one might also expect 
a similar behaviour from the solutions of equation~(\ref{sde}). In
fact, this is what happens when (\ref{sde}) is solved  using the
running coupling~(\ref{KTW}), which is also singular in the infrared
limit~\cite{KTW}.

     In order to obtain the low-energy asymptotic behaviour of the
dynamical mass function ${\mathcal M}(x)$, it is convenient to cast
the integral equation~(\ref{sde}) into a differential form, by
differentiating both sides   twice with respect to $x$ (see
also~\cite{KTW,atk})
\begin{equation}
\frac{d}{dx}\left[\frac{\frac{d}{dx}
\left[\frac{x{\mathcal M}(x)}{\alpha(x)}\right]}
{\frac{d}{dx}\left[\frac{x}{\alpha(x)}\right]}\right] = 
-\frac{1}{\pi} \frac{\alpha(x){\mathcal M}(x)}{x+{\mathcal M}^2(x)}.
\label{dif}
\end{equation}
By making use of the explicit form of the one-loop analytic invariant
charge~(\ref{AIC1L}) one can reduce this equation in the limit
$x\to0$ to
\begin{equation}
{\mathcal M}(v)
\left(\frac{d^2}{d v^2} + 2 \frac{d}{d v}\right) {\mathcal M}(v) =
\frac{8}{\beta_{0} v}\, , \qquad v \to -\infty,
\label{ndif}
\end{equation}
where $v = \ln x$. In deriving (\ref{ndif}) we have neglected the
factor  $z$ in the numerator of (\ref{AIC1L}), since in the limit
considered it is subleading next to~$1$. The above equation can be
solved analytically through successive iterations, giving rise, as
expected, to divergent solutions, which  are formally expressed in
terms of powers of  $v$ and~$\ln v$. The first iteration,
corresponding to the  leading infrared behaviour of the dynamical
mass function ${\mathcal M}(x)$, is obtained by omitting the
second-order derivative in equation~(\ref{ndif}). Then, the solution
is 
\begin{equation}
\label{lead}
{\mathcal M}(x) \simeq \sqrt{\frac{8}{\beta_0}\ln |\ln x|}, 
\qquad x \to 0.
\end{equation}  
The divergence of the dynamical quark mass function ${\mathcal M}(x)$
in the low-energy domain $x \to 0$ can be interpreted as a hint  for
confinement, see also discussion in~\cite{KTW,JMC80}.

     It is worthwhile to emphasize again that the obtained 
behaviour~(\ref{lead}) is restricted to the deep infrared domain
(i.e., for values $p^2 \ll \Lambda^2)$. Subleading corrections to
this solution may be systematically obtained from 
equation~(\ref{ndif}); however, they are of the same order as the
terms neglected when arriving from (\ref{dif}) to (\ref{ndif}), and
are therefore of little usefulness. It is also interesting to
compare  equation~(\ref{lead}) to the corresponding  solution
obtained by making use of the running coupling~(\ref{KTW}). The 
leading low-energy behaviour in that case is instead  ${\mathcal
M}_{\mbox{\tiny KTW}}(x) \simeq \sqrt{2C|\ln x|/\pi}$, when $x \to 0$.
Thus, we infer that for the case of the analytic invariant
charge~(\ref{AIC1L}) the infrared singularity of the dynamical mass
function ${\mathcal M}(x)$ is much milder than in the case of the
running coupling~(\ref{KTW}).

     For the sake of completeness, we finish this Section by
reporting the ultraviolet asymptotic expression for the dynamical
quark mass ${\mathcal M}(x)$. Since we are considering the case of
exact chiral symmetry (i.e., no bare mass, $m_0=0$), the conservation
of the axial-vector current eventually leads, for sufficient large
momenta, to   
\begin{equation}
{\mathcal M}(x) \simeq \frac{D}{x}(\ln x)^{\lambda -1}, \qquad
x \to \infty,
\end{equation}
where $\lambda = 12/(33-2\nf)$ is the anomalous dimension of the
mass, and $D$ is a constant independent of the renormalization point
and directly related to the quark condensate~\cite{miransky,
politzer}.

\section{Numerical solution for the mass function}
\label{Sect:NumSol}

\begin{figure}[t]
\begin{center}
\includegraphics[width=105mm]{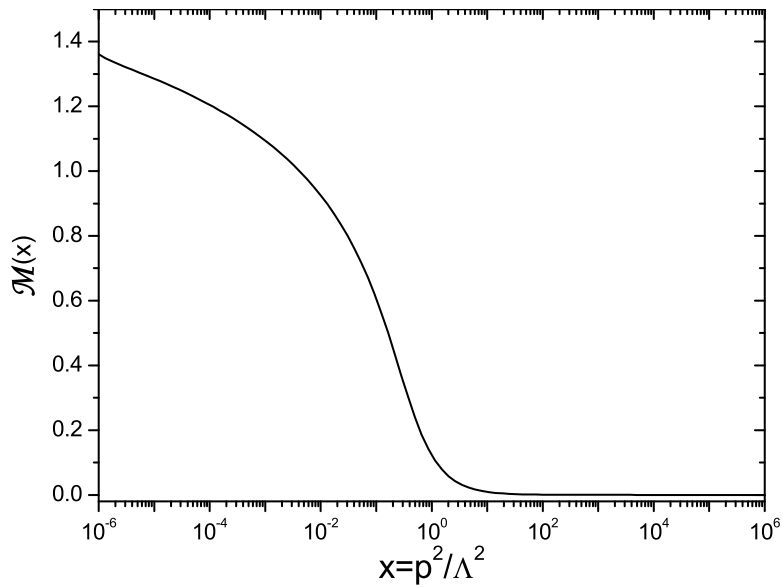}
\end{center}
\caption{The dimensionless quark dynamical mass ${\mathcal M}(x) =
M(x\Lambda^2)/\Lambda$, obtained from equation~(\ref{sde}) using the
one-loop analytic invariant charge~(\ref{AIC1L}) for $\nf=4$ quark
flavours.}
\label{csb-1aic}
\end{figure}

     The numerical solution for the dynamical quark mass function
${\mathcal M}(x)$, obtained solving directly the integral
equation~(\ref{sde}), is presented in Figure~\ref{csb-1aic}. Indeed,
one can see numerically a soft increase of the mass function
${\mathcal M}(x)$ when $x \to 0$, as suggested by the leading
behaviour~(\ref{lead}) extracted from the differential equation
analysis.  Regarding the ultraviolet region, we note that choosing
sufficiently  large values for the ultraviolet cutoff, the results
obtained are independent of the latter, since the integrals in
equation~(\ref{sde}) are ultraviolet convergent. Typically, our
solutions are evaluated within a momenta window of sixteen orders of
magnitude  $(10^{-8} \le x \le 10^{8})$,  which is sufficient to
ensure their stability.

     Now, with the solution for the dynamical mass at hand, we can
relate the value of the QCD scale parameter~$\Lambda$ to the pion
decay constant $f_{\pi}$, defined as the axial-vector transition
amplitude for an on-shell pion. This can be accomplished by making
use of the method developed by Pagels, Stokar~\cite{Pagels:1979hd},
and Cornwall \cite{JMC80}, which is a generalization of the
Goldberger--Treiman relation when  the momentum carried by the pion
is different from zero:
\begin{equation}
f_{\pi}^2= \frac{3\Lambda^2}{4\pi^2} \int_0^{\infty} 
\frac{y{\mathcal M}(y)}{\left[y+{\mathcal M}^2(y)\right]^2}
\left[{\mathcal M}(y) - 
\frac{y}{2}\frac{d{\mathcal M}(y)}{dy}\right] dy.
\label{fpim}
\end{equation}
Thus, one is able to fix the value for $\Lambda$, the only adjustable
parameter in this analysis, by requiring that the pion decay
constant   should assume its measured value of
$f_{\pi}=93\,$MeV~\cite{PDG}.   For the more favorable case of
$\nf=4$ active quarks   this procedure results in the
estimate~$\Lambda=880\,$MeV. 

     The inclusion of the higher loop corrections to the analytic 
running coupling $\al{}{an}(k^2)$ (see equation~(\ref{AICHLKL})) is 
not expected to alter the qualitative picture obtained above. This is
so because the most intrinsic feature of this charge, namely the 
infrared enhancement, persists after the incorporation of the loop
corrections; however, the type of singularity displayed at the origin
becomes slightly milder than in the one-loop case. Therefore, we
anticipate that the confining behaviour of ${\mathcal M}(x)$ will
also persist, but with a weaker infrared singularity then that of
equation~(\ref{lead}). On the other hand, in this case one would
expect a higher estimate for $\Lambda$, given that the loop
corrections lower the value of $\al{}{an}(k^2)$ in the entire range
of momenta~\cite{Review}. This fact, in turn, will lead to smaller
values for ${\mathcal M}(x)$, making  the saturation of the right
hand-side of (\ref{fpim}) more difficult, and forcing $\Lambda$ to
assume even higher values.

     We finish this Section by commenting on the veracity of the 
angular approximation given in equation~(\ref{angle}), and the
dependence of the  obtained results on it. It would be  certainly of
interest to establish whether the divergent nature of the solutions
persists, or is an artefact of the aforementioned approximation.
Indeed, one could envisage the possibility that the simultaneous
solution of the system~(\ref{aprop}) and~(\ref{bprop}) might actually
lead to a finite expression for  ${\mathcal M}(x)$, as $x\to 0$,
despite the fact that the kernel is divergent at the origin, due to
the enhanced form of the running charge~(\ref{AIC1L}) employed. In
order to address this point in some detail, we have not resorted to
the approximation of equation~(\ref{angle}), but have instead
performed the angular integration numerically, and subsequently
attempted to  solve the resulting coupled system. The numerical
integration of the final (momentum) integrals requires the
introduction of an infrared regulator~\cite{krein}; we have regulated
the kernels by carrying out the  replacement   $\alpha(k^2)/k^2
\rightarrow \alpha(k^2)/(k^2 + \mu^2)$. Evidently,  the solutions 
for the dynamical mass function depend now explicitly on the
regulator  $\mu$, and one should study the behaviour of ${\mathcal
M}(x,\mu^2)$  in the limit $\mu^2\to 0$. Our numerical analysis
revealed that there  is a critical value of $\mu^2$, of about $\mu^2
\simeq 10^{-7}\Lambda^2$,  below which no convergent solutions to 
the system~(\ref{aprop}) and~(\ref{bprop}) may be found. Although no
inescapable conclusion may be drawn from this fact, we interpret this
breakdown as a strong indication that the resulting solutions diverge
as $\mu^2\to 0$, in qualitative agreement with what was found when  
the angular approximation of equation~(\ref{angle}) was used.

\section{Conclusions}
\label{Sect:Concl}

     In this paper we have studied the compatibility between  the
infrared enhanced analytic charge of QCD  and chiral symmetry
breaking, through the detailed analysis  of a standard form of the
gap equation for the  quark propagator, where this former coupling
was incorporated. It turned out that, due to the  infrared
enhancement of the running coupling employed, the solutions  found
for the dynamical quark mass function~$M(p^2)$ were infrared
divergent. Following standard  arguments we have interpreted this
divergent behaviour as an indication of confinement. The final upshot
of this analysis was that the aforementioned analytic charge  is able
to break the chiral symmetry,  furnishing a reasonable ratio between
the QCD scale parameter~$\Lambda$ and the pion decay
constant~$f_{\pi}$.

     To be sure, the value of 880 MeV obtained for $\Lambda$  (with
four active quark flavours) appears elevated when compared to the
``standard'' value of $\Lambda$ of about 350 MeV quoted in the
literature  (see \cite{PDG} and references therein), but also when
compared  to the values for $\Lambda$ obtained within the analytic
approach itself  by resorting to different methods~\cite{Review},  
ranging between 500-600$\,$MeV.\footnote{It is worth noting that a  
number of authors have obtained values of $\Lambda$ in a similar
range from the study of the static quark-antiquark potential (see,
e.g.,~\cite{VScheme, FoLiWi79}).} We emphasize, however, that the
main purpose of the analysis presented is not so much to extract an
accurate value for  $\Lambda$, but rather check to what extend two
{\it a priori} different methods, the analytic approach and the   SD
equations, may coexist in a complementary and qualitatively
consistent picture. In that sense we consider the outcome of the
present work encouraging, especially when taking into account the
theoretical uncertainties intrinsic to both methods, and the  fact
that, unlike the majority of existing models, in our case $\Lambda$
is the only adjustable parameter available. 

     It would certainly be worthwhile attempting to improve the 
above picture by incorporating into the spectral density defining the
analytic charge (see equation~(\ref{AICHLKL})) contributions  from
nonperturbative effects (e.g., the operator product expansion (see
also~\cite{Maxwell})  and the nonlocal chiral quark
model~\cite{Dorokhov}). We hope to  be able to report progress in
this direction in the near future.

\ack

     The authors thank A.~Santamaria for useful discussions. This
work was supported by grants  SB2003-0065 of the Spanish Ministry of
Education, CICYT  FPA20002-00612, RFBR 05-01-00992, NS-2339.2003.2,
and by Coordena\c{c}\~{a}o de Aperfei\c{c}oamento de Pessoal de
N\'{\i}vel Superior (Capes/Brazil) through grant 2557/03-7 (A.C.A).

\end{document}